\documentclass[journal,article,accept,moreauthors,dvipdfm,12pt,a4paper]{mdpi} 
\lastpage{303}
\doinum{10.3390/e12030289}
\pubvolume{12}
\pubyear{2010}
\history{Received: 18 January 2010; in revised form: 16 February 2010 / Accepted: 23 February 2010 / Published: 26 February 2010}

\usepackage{amssymb,amsmath}
\usepackage{hyphenat}
\usepackage[sort&compress]{natbib}
\usepackage{hypernat}
\usepackage{caption}
\Title{Measurement Invariance, Entropy, and Probability}

\Author{Steven A.~Frank $^{1,2,\star}$ and D.~Eric Smith $^{2}$}

\address{%
$^{1}$ Department of Ecology and Evolutionary Biology, University of California, Irvine, \linebreak CA 92697--2525, USA\\
$^{2}$ Santa Fe Institute, 1399 Hyde Park Road, Santa Fe, NM 87501, USA
}

\corres{E-mail: safrank@uci.edu.}

\abstract{We show that the natural scaling of measurement for a
particular problem defines the most likely probability distribution
of observations taken from that measurement scale. Our approach
extends the method of maximum entropy to use measurement scale as a
type of information constraint.  We argue that a very common
measurement scale is linear at small magnitudes grading into
logarithmic at large magnitudes, leading to observations that often
follow Student's probability distribution which has a Gaussian shape
for small fluctuations from the mean and a power law shape for large
fluctuations from the mean. An inverse scaling often arises in which
measures naturally grade from logarithmic to linear as one moves
from small to large magnitudes, leading to observations that often
follow a gamma probability distribution.  A gamma distribution has a
power law shape for small magnitudes and an exponential shape for
large magnitudes. The two measurement scales are natural inverses
connected by the Laplace integral transform.  This inversion
connects the two major scaling patterns commonly found in nature. We
also show that superstatistics is a special case of an integral
transform, and thus can be understood as a particular way in which
to change the scale of measurement. Incorporating information about
measurement scale into maximum entropy provides a general approach
to the relations between measurement, information and probability.}


\newcommand{\renyi}{R{\'e}nyi}
\newcommand{\En}{{\cal{E}}}

\newcommand{\Tr}{\textrm{T}}
\newcommand{\GR}{\textrm{G}}
\newcommand{\Tri}{\Tr^{-1}}
\newcommand{\Trt}{\widetilde{\Tr}}
\newcommand{\ovra}{\ovr{1}{a}}

\newcommand{\Ga}{\alpha}
\newcommand{\Gb}{\beta}
\newcommand{\Gd}{\delta}

\newcommand{\Gg}{\gamma}

\newcommand{\Gl}{\lambda}
\newcommand{\GL}{\Lambda}

\newcommand{\Gs}{\sigma}

\newcommand{\angb}[1]{\left\langle#1\right\rangle}

\newcommand{\dd}{{\hbox{\rm d}}}

\newcommand{\Eq}[1]{Equation~(\ref{eq:#1})}

\newcommand{\ovr}[2]{{{#1}\over{#2}}}

\newcommand{\prt}{\partial}
\newcommand{\povr}[2]{\ovr{\prt #1}{\prt #2}}

\makeatletter
\def\NAT@penalty{\penalty\@m}%
\makeatother

\begin{document}

\section{Introduction}

Suppose you have a ruler that is about the length of your hand. With
that ruler, you can measure the size of all the visible objects in
your office.  That scaling of objects in your office with the length
of the ruler means that those objects have a natural linear scaling
in relation to your ruler.

Now consider the distances from your office to various galaxies.
Your ruler is of no use, because you cannot distinguish whether a
particular galaxy moves farther away by one ruler unit.  Instead,
for two galaxies, you can measure the ratio of distances from your
office to each galaxy.  You might, for example, find that one galaxy
is twice as far as another, or, in general, that a galaxy is some
percentage farther away than another.

Percentage changes define a ratio scale of measure, which has
natural units in logarithmic measure \citep{Hand04Measurement}. For
example, a doubling of distance always adds $\log(2)$ to the
logarithm of the distance, no matter what the initial distance.

Measurement naturally grades from linear at local magnitudes to
logarithmic at distant magnitudes when compared to some local
reference scale.  The transition between linear and logarithmic
varies between problems.  Measures from some phenomena remain
primarily in the linear domain, such as measures of height and
weight in humans. Measures for other phenomena remain primarily in
the logarithmic domain, such as cosmological distances.  Other
phenomena scale between the linear and logarithmic domains, such as
fluctuations in the price of financial assets
\citep{Aparicio01Empirical} or the distribution of income and wealth
\citep{dragulescu01exponential}.

The second section of this article shows how the characteristic
scaling of measurement constrains the most likely probability
distribution of observations.  We use the standard method of maximum
entropy to find the most likely probability distribution
\citep{Jaynes03Science}.  But, rather than follow the traditional
approach of starting with the information in the summary statistics
of observations, such as the mean or variance, we begin with the
information in the characteristic scale of measurement.  We argue
that measurement sets the fundamental nature of information and
shapes the probability distribution of observations. We present a
novel extension of the method of maximum entropy to incorporate
information about the scale of measurement.

The third section emphasizes the naturalness of the measurement
scale that grades from linear at small magnitudes to logarithmic at
large magnitudes.  This linear to logarithmic scaling leads to
observations that often follow a linear-log exponential or Student's
probability distribution. A linear-log exponential distribution is
an exponential shape for small magnitudes and a power law shape for
large magnitudes. Student's distribution is a Gaussian shape for
small fluctuations from the mean and a power law for large
fluctuations from the mean. The shapes correspond to linear scaling
at small magnitudes and logarithmic scaling at large magnitudes.
Many naturally observed patterns follow these distributions.  The
particular form depends on whether the measurement scale for a
problem is primarily linear, primarily logarithmic, or grades from
linear to logarithmic.

The fourth section inverts the natural linear to logarithmic scaling
for magnitudes.  Because magnitudes often scale from linear to
logarithmic as one moves from small to large magnitudes, inverse
measures often scale from logarithmic to linear as one moves from
small to large magnitudes.  This logarithmic to linear scaling leads
to observations that often follow a gamma probability distribution.
A gamma distribution is a power law shape for small magnitudes and
an exponential shape for large magnitudes, corresponding to
logarithmic scaling at small values and linear scaling at large
values.  The gamma distribution includes as special cases the
exponential distribution, the power law distribution, and the
chi-square distribution, subsuming many commonly observed patterns.

The fifth section demonstrates that the Laplace integral transform
provides the formal connection between the inverse measurement
scales. The Laplace transform, like its analytic continuation the
Fourier transform, changes a magnitude with dimension $d$ on one
scale into an inverse magnitude with dimension $1/d$ on the other
scale.  This inversion explains the close association between the
linear to logarithmic scaling as magnitudes increase and the inverse
scale that grades from logarithmic to linear as magnitudes increase.
We discuss the general role of integral transforms in changing the
scale of measurement.  Superstatistics is the averaging of a
probability distribution with a variable parameter over a
probability distribution for the variable parameter
\citep{Beck03Superstatistics}.  We show that superstatistics is a
special case of an integral transform, and thus can be understood as
a particular way in which to change the scale of measurement.

In the sixth section, we relate our study of measurement invariance
for continuous variables to previous methods of maximum entropy for
discrete variables. We also distinguish the general definition of
measurement scale by information invariance from our particular
argument about the commonness of linear-log scales.

In the discussion, we contrast our emphasis on the primacy of
measurement with alternative approaches to understanding
measurement, randomness, and probability.  One common approach
changes the definition of randomness and entropy to incorporate a
change in measurement scale \citep{Tsallis09Introduction}.  We argue
that our method makes more sense, because we directly incorporate
the change in measurement scale as a kind of information, rather
than alter the definition of randomness and entropy to match each
change in measurement scale.  It is measurement that changes
empirically between problems rather than the abstract meaning of
randomness and information. Although we focus on the duality between
linear to logarithmic scaling and its inverse logarithmic to linear
scaling, our general approach applies to any type of measure
invariance and measurement scale.

\section{Measurement, Information Invariance, and Probability}

We derive most likely probability distributions.  Our method follows
the maximum entropy approach
\citep{Jaynes57Information,Jaynes57Informationb,Jaynes03Science}.
That approach assumes that the most likely distribution has the
maximum amount of randomness, or entropy, subject to the constraint
that the distribution must capture all of the information available
to us.  For example, if we know the average value of a sample of
observations, and we know that all values from the underlying
probability distribution are positive, then all candidate
probability distributions must have only positive values and have a
mean value that agrees with the average of the empirically observed
values.  By maximum entropy, the most random distribution
constrained to have positive values and a fixed mean is the
exponential distribution.

We express the available information by constraints.  Typical
constraints include the average or variance of observations.  But we
must use all available information, which may include information
about the scale of measurement itself.  Previous studies have
discussed how the scale of measurement provides information.
However, that aspect of maximum entropy has not been fully developed
\citep{Jaynes03Science,Frank09The-common}.  Our goal is to develop
the central role of measurement scaling in shaping the commonly
observed \linebreak probability distributions.

In the following sections, we show how to use information about
measurement invariances and associated measurement scales to find
most likely probability distributions.

\subsection{Maximum entropy}

The method of maximum entropy defines the most likely probability
distribution as the distribution that maximizes a measure of entropy
(randomness) subject to various information constraints.  We write
the quantity to be maximized as
\begin{equation}\label{eq:maxEnt}
    \GL = \En - \Ga C_0 - \sum_{i=1}^n\Gl_iC_i
\end{equation}
where $\En$ measures entropy, the $C_i$ are the constraints to be
satisfied, and $\Ga$ and the $\Gl_i$ are the Lagrange multipliers to
be found by satisfying the constraints.  Let $C_0=\int p_y\dd y -1$
be the constraint that the probabilities must total one, where $p_y$
is the probability distribution function of $y$.  The other
constraints are usually written as $C_i= \int p_yf_i(y)\dd y
-\angb{f_i(y)}$, where the $f_i(y)$ are various transformed
measurements of $y$. Angle brackets denote mean values. A mean value
is either the average of some function applied to each of a sample
of observed values, or an a priori assumption about the average
value of some function with respect to a candidate set of
probability laws. If $f_i(y)=y^i$, then $\angb{y^i}$ are the moments
of the distribution---either the moments estimated from observations
or a priori values of the moments set by assumption.  The moments
are often regarded as ``normal'' constraints, although from a
mathematical point of view, any properly formed constraint can be
used.

Here, we confine ourselves to a single constraint of measurement. We
express that constraint with a more general notation, $C_1= \int
p_y\Tr[f(y)]\dd y -\angb{\Tr[f(y)]}$, where $\Tr()$ is a
transformation.  We could, of course, express the constraining
function for $y$ directly through $f(y)$.  However, we wish to
distinguish between an initial function $f(y)$ that can be regarded
as a normal measurement, in any sense in which one chooses to
interpret the meaning of normal, and a transformation of normal
measurements denoted by $\Tr()$ that arises from information about
the measurement scale.

The maximum entropy distribution is obtained by solving the set of equations
\begin{equation}\label{eq:maxEntSoln}
    \povr{\GL}{p_y} = \povr{\En}{p_y} - \Ga -
    \Gl\Tr[f(y)]=0
\end{equation}
where one checks the candidate solution for a maximum and obtains
$\Ga$ and $\Gl$ by satisfying the constraint on total probability
and the constraint on $\angb{\Tr[f(y)]}$. We assume that we can
treat entropy measures as the continuous limit of the discrete case.

In the standard approach, we define entropy by Shannon information
\begin{equation}\label{eq:shannonDef}
    \En=-\int p_y\log(p_y)\dd y
\end{equation}
which yields the solution of \Eq{maxEntSoln} as
\begin{equation}\label{eq:shannonSoln}
    p_y = ke^{- \Gl\Tr[f(y)]}
\end{equation}
where $k$ and $\Gl$ satisfy the two constraints.

\subsection{Measurement and transformation}\label{sectionInvar}

Maximum entropy, in order to be a useful method, must capture all of
the available information about a particular problem.  One form of
information concerns transformations to the measurement scale that
leave the most likely probability distribution unchanged.  Suppose,
for example, that we obtain the same information from measurements
of $x$ and transformed measurements, $\GR(x)$.  Put another way, if
one has access only to measurements on the $\GR(x)$ scale, one has
the same information that would be obtained if the measurements were
reported on the $x$ scale.  We say that the measurements $x$ and
$\GR(x)$ are equivalent with respect to information, or that the
transformation $x \rightarrow \GR(x)$ is an invariance
\citep{Hand04Measurement,luce08measurement,narens08meaningfulness}.

To capture this information invariance in maximum entropy, we must
express our measurements on a transformed scale.  In particular, we
must choose the transformation, $\Tr()$, for expressing measurements
so that
\begin{equation}\label{eq:transDef}
  \Tr(x) = \Gg + \Gd\Tr[\GR(x)]
\end{equation}
for some arbitrary constants $\Gg$ and $\Gd$.  Putting this
definition of $\Tr(x)$ into \Eq{shannonSoln} shows that we get the
same maximum entropy solution whether we use the direct scale $x$ or
the alternative measurement scale, $\GR(x)$, because the $k$ and
$\Gl$ constants will adjust to the constants $\Gg$ and $\Gd$ so that
the distribution remains unchanged.

Given the transformation $\Tr(x)$, the derivative of that
transformation expresses the information invariance in terms of
measurement invariance.  In particular, we have the following
invariance of the measurement scale under a change $\dd x$
\begin{equation}\label{eq:scalePropto}
  \dd\Tr(x) \propto \dd\Tr[\GR(x)]
\end{equation}
We may also examine $m_x=\Tr'(x)=\dd\Tr(x)/\dd x$ to obtain the
change in measurement scale required to preserve the information
invariance between $x$ and $\GR(x)$.

If we know the measurement invariance, $\GR(x)$, we can find the
correct transformation from \Eq{transDef}.  If we know the
transformation $\Tr(x)$, we can find $\GR(x)$ by inverting
\Eq{transDef} to obtain
\begin{equation}\label{eq:GfromT}
  \GR(x) = \Tri\left[\ovr{\Tr(x)-\Gg}{\Gd}\right]
\end{equation}
Alternatively, we may deduce the transformation $\Tr(x)$ by
examining the form of a given probability distribution and using
\Eq{shannonSoln} to find the associated transformation.

In summary, $x$ and $\GR(x)$ provide invariant information, and the
transformation of measurements $\Tr(x)$ captures that information
invariance in terms of measurement invariance.

\subsection{Example: ratio and scale invariance}

Suppose the information we obtain from positive-valued measurements
depends only on the ratio of measurements, $y_2/y_1$.  In this
particular case, all measurements with the same ratio map to the
same value, so we say that the measurement scale has ratio
invariance.  Pure ratio measurements also have scale invariance,
because ratios do not depend on the magnitude or scale of the
observations.

We express the invariances that characterize a measurement scale by
the transformations that leave the information in the measurements
unchanged
\citep{Hand04Measurement,luce08measurement,narens08meaningfulness}.
If we obtain values $x$ and use the measurement scale from the
transformation $\Tr(x) = \log(x)$, the information in $x$ is the
same as in $\GR(x) = x^c$, because $\Tr(x) = \log(x)$ and
$\Tr[\GR(x)] = c\log(x)$, so in general $\Tr(x) \propto
\Tr[\GR(x)]$, which means that the information in the measurement
scale given by $\Tr(x)$ is invariant under the transformation
$\GR(x)$.

We can express the invariance in a way that captures how measurement
relates to information and probability.  The transformation $\Tr(x)
= \log(x)$ shrinks increments on the uniform scaling of $x$ so that
each equally spaced increment on the original uniform scale shrinks
to length $1/x$ on the transformed scale.  We can in general
quantify the deformation in incremental scaling by the derivative of
the transformation $\Tr(x)$ with respect to $x$.  In the case of the
logarithmic measurement scale with ratio invariance, the measure
invariance in \Eq{scalePropto} is
\begin{equation*}
  \dd\log(x)\propto\dd\log[\GR(x)] \Rightarrow \ovr{1}{x} \propto \ovr{c}{x}
\end{equation*}
showing in another way that the logarithmic measure $\Tr(x)$ is
invariant under the transformation $\GR(x)$.  With regard to
probability or information, we can think of the logarithmic scale
with ratio invariance as having an expected density of probability
per increment in proportion to $1/x$, so that the expected density
of observations at scale $x$ decreases in proportion to $1/x$.
Roughly, we may also say that the information value of an increment
decreases in proportion to $1/x$.  For example, the increment length
of our hand is an informative measure for the visible objects near
us, but provides essentially no information on a cosmological scale.

If we have measurements $f(y) = y$, and we transform those
measurements in a way consistent with a ratio and scale invariance
of information, then we have the transformed measures $\Tr[f(y)] =
\log(y)$.  The constraint for maximum entropy corresponds to
$\angb{\log(y)}$, which is logarithm of the geometric mean of the
observations on the direct scale $y$.  Given that constraint, the
maximum entropy distribution is a power law
\begin{equation*}
    p_y=ke^{-\Gl\Tr[f(y)]}=ke^{-\Gl\log(y)}=ky^{-\Gl}
\end{equation*}
For $y\ge 1$, we can solve for the constants $k$ and $\Gl$, yielding
$p_y= \Gd y^{-(1+\Gd)}$, with $\Gd=1/\angb{\log(y)}$.

\section{The linear to Logarithmic Measurement Scale}

\subsection{Measurement}

In the previous section, we obtained ratio and scale invariance with
a measure $m_x=\Tr'(x) \propto 1/x$.  In this section, we consider
the more general measure
\begin{equation*}
  m_x \propto \ovr{1}{1+bx}
\end{equation*}
At small values of $x$, the measure becomes linear, $m_x \propto 1$,
and at large values of $x$, the measure becomes ratio invariant
(logarithmic), $m_x \propto 1/x$.  This measure has scale dependence
with ratio invariance at large scales, because the measure changes
with the magnitude (scale) of $x$, becoming ratio invariant at large
values of $x$.  The parameter $b$ controls the scale at which the
measure grades between linear and logarithmic.

Given $m_x=\Tr'(x)$, we can integrate this deformation of
measurement to obtain the associated scale of measurement as
\begin{equation}\label{eq:linearlog}
  \Tr(x) = \ovr{1}{a}\log(1+bx)=\log(1+bx)^\ovr{1}{a}\propto \log(1+bx)
\end{equation}
where we have expressed the proportionality constant as $1/a$ and we
have dropped the constant of integration. The expression
$\log(1+bx)$ is just a logarithmic measurement scale for positive
values in relation to a fixed origin at $x=0$, because $\log(1) =
0$.  The standard logarithmic expression, $\log(x)$, has an implicit
origin for positive values at $x=1$, which is only appropriate for
purely ratio invariant problems with no notion of an origin to set
the scale of magnitudes.  In most empirical problems, there is some
information about the scaling of magnitudes.  Thus, $\log(1+bx)$ is
more often the natural measurement scale.

Next, we seek an expression $\GR(x)$ to describe the information
invariance in the measurement scale, such that the information in
$x$ and in $\GR(x)$ is the same.  The expression in
\Eq{scalePropto}, \linebreak $\dd\Tr(x)\propto\dd \GR[\Tr(x)]$, sets the
condition for information invariance, leading to
\begin{equation}\label{eq:Gtrans1}
  \GR(x)=\ovr{(1+bx)^\ovra - 1}{b}
\end{equation}
On the measurement scale $\Tr(x)$, the information in $x$ is the same as in $\GR(x)$, because
\begin{equation*}
  \dd\Tr(x)\propto\dd\Tr[\GR(x)] \Rightarrow \ovr{b/a}{1+bx} \propto \ovr{b/a^2}{1+bx}
\end{equation*}

We now use $x=f(y)$ to account for initial normal measures that may
be taken in any way we choose.  Typically, we use direct values,
$f(y)=y$, or squared values, $f(y)=y^2$, corresponding to initial
measures related to the first and second moments---the average and
variance.  For now, we use $f(y)$ to hold the place of whatever
direct values we will use.  Later, we consider the interpretations
of the first and second moments.

\subsection{Probability}

The constraint for maximum entropy corresponds to
$\angb{\Tr[f(y)]}=\angb{\log[1+bf(y)]^\ovra}$, a value that
approximately corresponds to an interpolation between the linear
mean and the geometric mean of $f(y)$.  Given that constraint, the
maximum entropy distribution from \Eq{shannonSoln} is
\begin{equation}\label{eq:linearLogResult}
    p_y  \propto [1+bf(y)]^{-\Ga}
\end{equation}
where $\Ga=\Gl/a$ acts as a single parameter chosen to satisfy the
constraint, and $b$ is a parameter derived from the measurement
invariance that expresses the natural scale of measurement for a particular problem.

From \Eq{linearLogResult}, we can express simple results when in
either the purely linear or purely logarithmic regime.  For small
values of $bf(y)$ we can write $p_y  \propto e^{-\Ga bf(y)}$. For
large values of $bf(y)$ we can write $p_y \propto f(y)^{-\Ga}$,
where we absorb $b^{-\Ga}$ into the proportionality constant.  Thus,
the probability distribution grades from exponential in $f(y)$ at
small magnitudes to a power law in $f(y)$ at large magnitudes,
corresponding to the grading of the linear to logarithmic
measurement scale.

\subsection{Transition between linear and logarithmic scales}

We mentioned that one can obtain the parameter $\Ga$ in
\Eq{linearLogResult} directly from the constraint
$\angb{\Tr[f(y)]}$, which can be calculated directly from observed
values of the process or set by assumption.  What about the
parameter $b$ that sets the grading between the linear and
logarithmic regimes?

When we are in the logarithmic regime at large values of $bf(y)$,
probabilities scale as $p_y \propto f(y)^{-\Ga}$ independently of
$b$.  Thus, with respect to $b$, we only need to know the magnitude
of observations above which ratio invariance and logarithmic scaling
become reasonable descriptions of the measurement scale.

In the linear regime, $p_y  \propto e^{-\Ga bf(y)}$, thus $b$ only
arises as a constant multiplier of $\Ga$ and so can be subsumed into
a single combined parameter $\Gb=\Ga b$ estimated from the single
constraint.  However, it is useful to consider the meaning of $b$ in
the linear regime to provide guidance for how to interpret $b$ in
the mixed regime in which we need the full expression in
\Eq{linearLogResult}.

When $f(y)=y$, the linear regime yields an exponential distribution
$p_y  \propto e^{-\Ga by}$.  In this case, $b$ weights the intensity
or rate of the process $\Ga$ that sets the scaling of the
distribution.

When $f(y)=y^2$, the linear regime yields a Gaussian distribution
$p_y  \propto e^{-\Ga by^2}$, where $2\Ga b$ is the reciprocal of
the variance that defines the precision of measurements---the amount
of information a measurement provides about the location of the
average value.  In this case, $b$ weights the precision of
measurement.  The greater the value of $b$, the more information per
increment on the measurement scale.

\subsection{Linear-log exponential distribution}

When $f(y)=y$, we obtain from \Eq{linearLogResult} what we will call
the linear-log exponential distribution
\begin{equation}\label{eq:linearlogExp}
    p_y  \propto [1+by]^{-\Ga}
\end{equation}
for $y>0$.  This distribution is often called the generalized type
II Pareto distribution or the Lomax distribution
\citep{Johnson94Distributions}. Small values of $by$ lead to an
exponential shape, $p_y \propto e^{-\Ga by}$.  Large values of $by$
lead to power law tails, $p_y \propto y^{-\Ga}$.  The parameter $b$
determines the grading from the exponential to the power law. Small
values of $b$ extend the exponential to higher values of $y$,
whereas large values of $b$ move the extent of the power law shape
toward smaller values of $y$.  Many natural phenomena follow a
linear-log exponential distribution \citep{Tsallis09Introduction}.

\subsection{Student's distribution}

When $f(y)=y^2$, we obtain from \Eq{linearLogResult} Student's distribution
\begin{equation}\label{eq:students}
    p_y  \propto [1+by^2]^{-\Ga}
\end{equation}
Here, we assume that $y$ expresses deviations from the average.
Small deviations lead to a Gaussian shape around the mean, $p_y
\propto e^{-\Ga by^2}$.  Large deviations lead to power law tails,
$p_y \propto f(y)^{-\Ga}$.  The parameter $b$ determines the grading
from the Gaussian to the power law. Small values of $b$ expand the
Gaussian shape far from the mean, whereas large values of $b$ move
the extent of the power law shape closer to the central value at the
average.  Many natural phenomena expressed as deviations from a
central value follow Student's distribution
\citep{Tsallis09Introduction}.

The ubiquity of both Student's distribution and the linear-log
exponential distribution arises from the fact that the grading
between linear measurement scaling at small magnitudes and
logarithmic measurement scaling at large magnitudes is inevitably
widespread.  Many cases will be primarily in the linear regime and
so be mostly exponential or Gaussian except in the extreme tails.
Many other cases will be primarily in the logarithmic regime and so
be mostly power law except in the regime of small deviations near
the origin or the central location. Other cases will produce
measurements across both scales and their transition.

\section{The Inverse Logarithmic to Linear Measurement Scale}

We have argued that the linear to logarithmic measurement scale is
likely to be common. Magnitudes such as time or distance naturally
grade from linear at small scales to logarithmic at large scales.

Many problems measure inverse dimensions, such as the reciprocals of
time or distance.  If magnitudes of time or space naturally grade
from linear to logarithmic as scale increases from small to large,
then how do the reciprocals scale?  In this section, we argue that
the inverse scale naturally grades from logarithmic to linear as
scale increases from small to large.

We first describe the logarithmic to linear measurement scale and
its consequences for probability.  We then show the sense in which
the logarithmic to linear scale is the natural inverse of the linear
to logarithmic scale.

\subsection{Measurement}

The transformation
\begin{equation*}
  \Tr(x) = x + b\log(x)
\end{equation*}
corresponds to the change in measurement scale $m_x=1+b/x$.  As $x$
becomes small, the measurement scaling $m_x\rightarrow1/x$ becomes
the ratio-invariant logarithmic scale.  As $x$ increases, the
measurement scaling $m_x\rightarrow1$ becomes the uniform measure
associated with the standard linear scale.  Thus, the scaling
$m_x=1+b/x$ interpolates between logarithmic and linear
measurements, with the weighting of the two scales shifting from
logarithmic to linear as $x$ increases from small to large values.

\subsection{Probability}

The constraint for maximum entropy corresponds to
$\angb{\Tr[f(y)]}=\angb{f(y)+b\log[f(y)]}$, a value that
interpolates between the linear mean and the geometric mean of
$f(y)$.  Given that constraint, the maximum entropy distribution is
$p_y  \propto f(y)^{-\Gl b}e^{-\Gl f(y)}$, with $\Gl$ chosen to
satisfy the constraint.

The direct measure $f(y)=y$ for positive values is the gamma distribution
\begin{equation}\label{eq:gamma}
    p_y  \propto y^{-\Gl b}e^{-\Gl y}
\end{equation}
As $y$ becomes small, the distribution approaches a power law form,
$p_y\propto y^{-\Gl b}$. As $y$ becomes large, the distribution
approaches an exponential form in the tails, $p_y\propto e^{-\Gl
y}$.  Thus, the distribution grades from power law at small scales
to exponential at large scales, corresponding to the measurement
scale that grades from logarithmic to linear as magnitude increases.
Larger values of $b$ extend the power law to higher magnitudes by
pushing the logarithmic to linear change in measure to higher
magnitudes.  The combination of power law and exponential shapes in
the gamma distribution is the direct inverse of the linear-log
exponential distribution given in \Eq{linearlogExp}.

The squared values $f(y)=y^2$, which we interpret as squared deviations from the average value, \linebreak lead to
\begin{equation}\label{eq:gammaGauss}
    p_y  \propto y^{-\Gl b}e^{-\Gl y^2 / 2}
\end{equation}
where the exponent of two on the first power law component is
subsumed in the other parameters.  This distribution is a power law
at small scales with Gaussian tails at large scales, providing the
inverse of Student's distribution in \Eq{students}. This
distribution is a form of the generalized gamma distribution
\citep{Johnson94Distributions}, which we call the gamma-Gauss
distribution.  This distribution may, for example, arise as the sum
of truncated power laws or L\'evy flights \citep{Frank09The-common}.

\section{Integral Transforms and Superstatistics}

The previous sections showed that linear to logarithmic scaling has
a simple relation to its inverse of logarithmic to linear scaling.
That simple relation suggests that the two inverse scales can be
connected by some sort of transformation of measure.  We will now
show the connection.

Suppose we start with a particular measurement scale given by
$\Tr(x)$ and its associated probability distribution given by
\begin{equation*}
    p_x \propto e^{-\Ga\Tr(x)}
\end{equation*}
Consider a second measurement scale $\Trt(\Gs)$ with associated probability distribution
\begin{equation*}
    p_\Gs \propto e^{-\Gl\Trt(\Gs)}
\end{equation*}
What sort of transformation relates the two measurement scales?

The integral transforms often provide a way to connect two measurement scales.  For example, we could write
\begin{equation}\label{eq:integralTransform}
    p_x \propto \int_{\Gs^-}^{\Gs^+}p_\Gs g_{x|\Gs}\dd\Gs
\end{equation}
This expression is called an integral transform of $p_\Gs$ with
respect to the transformation kernel $g_{x|\Gs}$.  If we interpret
$g_{x|\Gs}$ as a probability distribution of $x$ given a parameter
$\Gs$, and $p_\Gs$ as a probability distribution over the variable
parameter $\Gs$, then the expression for $p_x$ is called a
superstatistic: the probability distribution, $p_x$, that arises
when one starts with a different distribution, $g_{x|\Gs}$, and
averages that distribution over a variable parameter with
distribution $p_\Gs$ \citep{Beck03Superstatistics}.

It is often useful to think of a superstatistic as an integral
transform that transforms the measurement scale.  In particular, we
can expand \Eq{integralTransform} as
\begin{equation}\label{eq:integralTransform2}
    e^{-\Ga\Tr(x)} \propto \int_{\Gs^-}^{\Gs^+}e^{-\Gl\Trt(\Gs)} g_{x|\Gs}\dd\Gs
\end{equation}
which shows that the transformation kernel $g_{x|\Gs}$ changes the
measurement scale from $\Trt(\Gs)$ to $\Tr(x)$.  It is not necessary
to think of $g_{x|\Gs}$ as a probability distribution---the
essential role of $g_{x|\Gs}$ concerns a change in measurement
scale.

The Laplace transform provides the connection between our inverse
linear-logarithmic measurement scales.  To begin, expand the right
side of \Eq{integralTransform2} using the Laplace transform kernel
$g_{x|\Gs}=e^{-\Gs x}$, and use the inverse logarithmic to linear
measurement scale, $\Trt(\Gs)=\Gs+b\log(\Gs)$.  Integrating from
zero to infinity yields
\begin{equation*}
    e^{-\Ga\Tr(x)} \propto (1+x/\Gl)^{b\Gl-1}
\end{equation*}
with the requirement that $b\Gl<1$.  From this, we have $\Tr(x)
\propto \log(1+x/\Gl)$, which is the linear to logarithmic scale.
Thus, the Laplace transform inverts the logarithmic to linear scale
into the linear to logarithmic scale.  The inverse Laplace transform
converts in the other direction.

If we use $x=y$, then the transform relates the linear-log
exponential distribution of \Eq{linearlogExp} to the gamma
distribution of \Eq{gamma}.  If we use $x=y^2$, then the transform
relates Student's distribution of \Eq{students} to the gamma-Gauss
distribution of \Eq{gammaGauss}.

The Laplace transform inverts the measurement scales.  This
inversion is consistent with a common property of Laplace
transforms, in which the transform inverts a measure with dimension
$d$ to a measure with dimension $1/d$. One sometimes interprets the
inversion as a change from a direct measure to a rate or frequency.
Here, it is only the inversion of dimension that is significant. The
inversion arises because, in the transformation kernel
$g_{x|\Gs}=e^{-\Gs x}$, the exponent $\Gs x$ is typically
non-dimensional, so that the dimensions of $\Gs$ and $x$ are
reciprocals of each other.  The transformation takes a distribution
in $\Gs$ given by $p_\Gs$ and returns a distribution in $x$ given by
$p_x$. Thus, the transformation typically inverts \linebreak the dimension.

\section{Connections and Caveats}

\subsection{Discrete versus continuous variables}

We used measure invariance to analyze maximum entropy for continuous
variables.  We did not discuss discrete variables, because measure
invariance applied to discrete variables has been widely and
correctly used in maximum entropy
\citep{Jaynes03Science,Sivia06Tutorial,Frank09The-common}.  In the
Discussion, we describe why previous attempts to apply invariance to
continuous variables did not work in general.  That failure
motivated our \linebreak current study.

Here, we briefly review measure invariance in the discrete case for
comparison with our analysis of continuous variables.  We use the
particular example of $N$ Bernoulli trials with a sample measure of
the number of successes $y=0,1,\ldots,N$.  Frank
\citep{Frank09The-common} describes the measure invariance for this
case: ``How many different ways can we can obtain $y=0$ successes in
$N$ trials? Just one: a series of failures on every trial. How many
different ways can we obtain $y=1$ success? There are $N$ different
ways: a success on the first trial and failures on the others; a
success on the second trial, and failures on the others; and so on.
The uniform solution by maximum entropy tells us that each different
combination is equally likely. Because each value of $y$ maps to a
different number of combinations, we must make a correction for the
fact that measurements on $y$ are distinct from measurements on the
equally likely combinations. In particular, we must formulate a
measure\ldots that accounts for how the uniformly distributed basis
of combinations translates into variable values of the number of
successes, $y$. Put another way, $y$ is invariant to changes in the
order of outcomes given a fixed number of successes. That invariance
captures a lack of information that must be included in our
analysis.''

In this particular discrete case, transformations of order do not
change our information about the total number of successes.  Our
measurement scale expresses that invariance, and that invariance is
in turn captured in the maximum entropy distribution.

The nature of invariance is easy to see in the discrete case by
combinatorics. The difficulty in past work has been in figuring out
exactly how to capture the same notion of invariance in the
continuous case.  We showed that the answer is perhaps as simple as
it could be:  use the transformations that do not change information
in the context of a particular problem.  Jaynes
\citep{Jaynes03Science} hinted at this approach, but did not develop
and apply the idea in a general way.

\subsection{General measure invariance versus particular linear-log scales}

Our analysis followed two distinct lines of argument.  First, we
presented the general expression for invariance as a form of
information in maximum entropy. We developed that expression
particularly for the case of continuous variables.  The general
expression sets the conditions that define measurement scales and
the relation between measurement and probability.  But the general
expression does not tell us what particular measurement scale will
arise in any problem.

Our second line of argument claimed that various types of grading
between linear and logarithmic measures arise very commonly in
natural problems.  Our argument for commonness is primarily
inductive and partially subjective.  On the inductive side, the
associated probability distributions seem to be those most commonly
observed in nature. On the subjective side, the apparently simplest
assumptions about invariance lead to what we called the common
gradings between linear and logarithmic scales.    We do not know of
any way to prove commonness or naturalness.  For now, we are content
that the general mathematical arguments lead in a simple way to
those probability distributions that appear to arise commonly in
nature.

A different view of what is common or what is simple would of course
lead to different information invariances, measurement scales, and
probability distributions.  In that case, our general mathematical
methods would provide the tools by which to analyze the alternative
view.

\section{Discussion}

We developed four topics.  First, we provided a new extension to the
method of maximum entropy in which we use the measurement scale as a
primary type of information constraint.  Second, we argued that a
measurement scale that grades from linear to logarithmic as
magnitude increases is likely to be very common.  The linear-log
exponential and Student's distributions follow immediately from this
measurement scale.  Third, we showed that the inverse measure that
grades from logarithmic at small scales to linear at large scales
leads to the gamma and gamma-Gauss distributions.  Fourth, we
demonstrated that the two measurement scales are natural inverses
related by the Laplace integral transform.  Superstatistics are a
special case of integral transforms and can be understood as changes
in measurement scale.

In this discussion, we focus on measurement invariance, alternative
definitions of entropy, and maximum entropy methods.


Jaynes \citep{Jaynes03Science} summarized the problem of
incorporating measurement invariance as a form of information in
maximum entropy.  The standard conclusion is that one should use
relative entropy to account for measurement invariance.  In our
notation, for a measurement scale $\Tr(y)$ with measure deformation
\linebreak $m_y=\Tr'(y)$, the form of relative entropy is the Kullback-Leibler
divergence
\begin{equation*}
  \En = -\int p_y\log\left(\ovr{p_y}{m_y}\right)\dd y
\end{equation*}
in which the $m_y$ is proportional to a prior probability
distribution that incorporates the information from the measurement
scale and leads to the analysis of maximum relative entropy.  This
approach works in cases where the measure change, $m_y$, is directly
related to a change in the probability measure.  Such changes in
probability measure typically arise in combinatorial problems, such
as a type of measurement that cannot distinguish between the order
of elements in sets.

For continuous deformations of the measurement scale, using $m_y$ as
a relative scaling for probability does not always give the correct
answer.  In particular, if one uses the constraint $\angb{f(y)}$ and
the measure $m_y$ in the above definition of relative entropy, the
maximum relative entropy gives the \linebreak probability distribution
\begin{equation*}
  p_y \propto m_ye^{-\Gl f(y)}
\end{equation*}
which is often not the correct result.  Instead, the correct result
follows from the method we gave, in which the information from
measurement invariance is incorporated by transforming the
constraint as $\angb{\Tr[f(y)]}$, yielding the maximum entropy
solution
\begin{equation*}
  p_y \propto e^{-\Gl \Tr[f(y)]}
\end{equation*}

It is possible to change the definition of entropy, such that
maximum entropy applied to the transformed measure of entropy plus
the direct constraint $\angb{f(y)}$ gives the correct answer
\citep{Tsallis09Introduction}.  The resulting probability
distributions are of course the same when transforming the
constraint or using an appropriate matching transformation of the
entropy measure.  We discuss the mathematical relation between the
alternative transformations in a later paper.

We prefer the transformation of the constraint, because that
approach directly shows how information about measurement alters
scale and is incorporated into maximum entropy.  By contrast,
changing the definition of entropy requires each measurement scale
to have its own particular definition of entropy and information.
Measurement is inherently an empirical phenomenon that is particular
to each type of problem and so naturally should be changed as the
nature of the problem changes.  The abstract notions of entropy and
information are not inherently empirical factors that change among
problems, so it seems perverse to change the definition of
randomness with each transformation of measurement.

The Tsallis and \renyi\ entropy measures are transformations of
Shannon entropy that incorporate the scaling of measurement from
linear to logarithmic as magnitude increases
\citep{Tsallis09Introduction}.  Those forms of entropy therefore
could be used as a common alternative to Shannon entropy whenever
measurements scale linearly to logarithmically.  Although
mathematically correct, such entropies change the definition of
randomness to hide the particular underlying transformation of
measurement.  That approach makes it very difficult to understand
how alternative measurement scales alter the expected types of
\linebreak probability distributions.

\section{Conclusions}

Linear and logarithmic measurements seem to be the most common
natural scales.  However, as magnitudes change, measurement often
grades between the linear and logarithmic scales.  That transition
between scales is often overlooked.  We showed how a measurement
scale that grades from linear to logarithmic as magnitude increases
leads to some of the most common patterns of nature expressed as the
linear-log exponential distribution and Student's distribution.
Those distributions include the exponential, power law, and Gaussian
distributions as special cases, and also include hybrids between
those distribution that must commonly arise when measurements span
the linear and logarithmic regimes.

We showed that a measure grading from logarithmic to linear as
magnitude increases is a natural inverse scale.  That measurement
scale leads to the gamma and gamma-Gauss distributions.  Those
distributions are also composed of exponential, power law, and
Gaussian components.  However, those distributions have the power
law forms at small magnitudes corresponding to the logarithmic
measure at small magnitudes, whereas the inverse scale has the power
law components at large magnitudes corresponding to the logarithmic
measure at large magnitudes.

The two measurement scales are natural inverses connected by the
Laplace transform.  That transform inverts the dimension, so that a
magnitude of dimension $d$ on one scale becomes an inverse magnitude
of dimension $1/d$ on the other scale.  Inversion connects the two
major scaling patterns commonly found in nature.  Our methods of
incorporating information about measurement scale into maximum
entropy also apply to other forms of measurement scaling and
invariance, providing a general method to study the relations
between measurement, information, and probability.

\section*{Acknowledgements}

SAF is supported by National Science Foundation grant EF-0822399,
National Institute of General Medical Sciences MIDAS Program grant
U01-GM-76499, and a grant from the James S.~McDonnell Foundation.
DES thanks Insight Venture Partners for support.


\bibliographystyle{mdpi}
\makeatletter
\renewcommand\@biblabel[1]{#1. }
\makeatother

\end{document}